# Managing Smartphone Crowdsensing Campaigns through the Organicity Smart City Platform

**Dimitrios Amaxilatis**
**Evangelos Lagoudianakis**
Computer Technology Institute & Press "Diophantus", Patras, Greece
and
Computer Eng. & Informatics Dpt, University of Patras, Greece
amaxilat@cti.gr,
lagoudiana@ceid.upatras.gr

**Georgios Mylonas**
Computer Technology Institute & Press "Diophantus", Patras, Greece
mylonasg@cti.gr

**Evangelos Theodoridis**
Intel Labs Europe
London, UK
evangelos.theodoridis@intel.com

## Abstract
We briefly present the design and architecture of a system that aims to simplify the process of organizing, executing and administering crowdsensing campaigns in a smart city context over smartphones volunteered by citizens. We built our system on top of an Android app substrate on the end-user level, which enables us to utilize smartphone resources. Our system allows researchers and other developers to manage and distribute their "mini" smart city applications, gather data and publish their results through the Organicity smart city platform. We believe this is the first time such a tool is paired with a large scale IoT infrastructure, to enable truly city-scale IoT and smart city experimentation.

## Author Keywords
Crowdsensing; sensors; campaign; experimentation management; smart city; Android; smartphone; volunteers; experiment; IoT; mobile.

## ACM Classification Keywords
H.5.m. Information interfaces and presentation: Miscellaneous; H.3.4. Systems and Software: Distributed Systems.

**Organicity**

Organicity is an EU project that puts people at the centre of the development of future cities. The project brings together 3 leading smart cities: Aarhus, London and Santander and utilizes both static IoT infrastructure and mobile devices in order to provide a platform that enables citizens and developers to co-create smart city applications and solutions. In this context, it will allow for IoT experiments to be run through its platform. One category of such experiments involves the participation of volunteers through their smartphones, the organization of which is facilitated through the system presented here. Smartphones can contribute their own integrated sensor readings, or communicate wirelessly, e.g., over Bluetooth, with other IoT devices, e.g., Arduino sensing boards, thus acting as proxies. Our system helps to deploy and monitor the progress of such crowdsensing campaigns.

**Introduction and Related Work**

Smart cities are currently, together with the Internet of Things, one of the most promising research fields in informatics. Although both have been around for a number of years, only recently have the technological and financial conditions allowed for real large-scale IoT city deployments and system development. In this context, a number of research projects, like SmartSantander[1], focused on building large IoT infrastructures inside city centers and offering researchers and companies with a "testbed" to develop and test their systems and applications.

Meanwhile, current mainstream smartphones utilize a number of integrated sensors, while also having the necessary networking interfaces to communicate with IoT devices like Arduino, smartwatches or fitness trackers (i.e., Bluetooth LE, NFC, etc.). Also, the maker movement expressed in communities like hackerspaces and fab labs has surfaced in many cities around the world to activate citizens that cooperate in conjunction with existing activist and citizen groups in the creation of a more human centric environment in large metropolitan environments.

In light of these advancements, recent projects like Organicity[2] (OC) aim to combine the aforementioned approaches and co-create together with citizens, researchers and city authorities, new smart city solutions. Crowdsensing (i.e., tasking groups of volunteers to gather various kinds of data using IoT devices) is one of the directions taken to address this challenge, helping build a smart city data repository.

In this work we describe the design and architecture of a system aiming to simplify the management, deployment and execution of crowdsensing campaigns. Our work acts as a component of the large-scale Organicity smart city infrastructure. Overall, the system:

> Uses a smartphone experimentation substrate, which enables execution of Android "mini" applications, written as Android Java code.
> Enables the definition of crowdsensing campaigns in great detail, using spatiotemporal and other types of constraints.
> Enables the monitoring of the execution of such campaigns and their dynamic redefinition, i.e., change the constraints at any time.
> Works as a part of the Organicity platform, providing the option of making data available through its Urban Data Observatory[3].

Regarding related previous work, the smartphone experimentation substrate utilized here has been described in greater detail in [1]. Also, the area of crowdsensing is described extensively in [2], providing both the theoretical background and a review of a number of approaches currently utilized. Our work has similarities with the approaches and ideas described in [3,4,5,6]; however, apart from aiming at providing a pragmatic solution to the crowdsensing problem, it is also integrated with a smart city platform, allowing end-users to benefit from this interoperability in various ways (data storage, visualization, interfacing to other systems, community management etc.). We also allow easy interconnection to other IoT devices for scenarios like the one described in [7].

---

[1] http://smartsantander.eu

[2] http://organicity.eu

[3] http://organicity.smartcitizen.me/

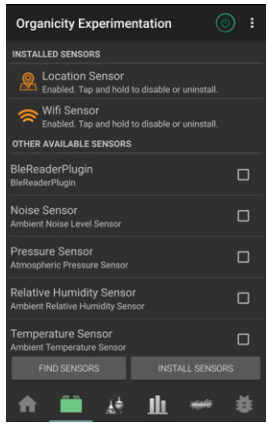

Figure 1: Smartphone Experimentation Android app, available Sensors view.

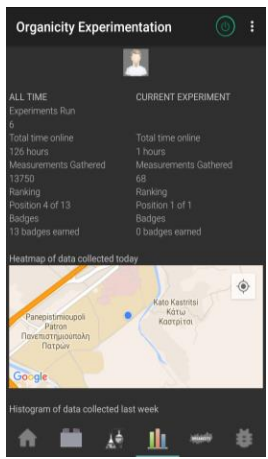

Figure 2: Smartphone Experimentation Android app, data collection statistics view.

## System Overview and Architecture

Following the dominant concept in IoT experimentation testbeds, smartphones execute "experiments", which produce readings which are then made available through Organicity's Urban Data Observatory. Overall, citizens perform the experiments during their commute or leisure activities transparently; the overall idea is to utilize their smartphones in order to execute IoT experiments on devices other than the existing Organicity infrastructure. There are two discrete Organicity components utilized:

> A server/portal component dedicated to submitting, monitoring and managing the execution of IoT experiments, described here.
> A smartphone component (see [1]) dedicated to actually running the IoT experiments on smartphones and communicating the results to the server component.

The IoT experimenters essentially utilize the sensors and networking interfaces of a smartphone carried by volunteers, which are then reported to the UDO as parts of the Organicity IoT infrastructure. The whole process of deploying and running the IoT experiments on smartphones, along with communicating back to Organicity is *transparent* to IoT experimenters.

The Smartphone Experimentation component is based on executing OSGi plugins as part of a main experimentation application; in other words, *the functionality of the main smartphone app is dynamically updated/ augmented* by downloading and executing such plugins. The tool is based on the use of OSGi plugins by Ambient Dynamix[4]. Android OSGi plugins allow us to run different experiments on demand using a single application. The execution of such experiments is also dependent on the OC participants' preferences, i.e., in terms of sensing modalities that they wish to contribute to the project, thus greatly simplifying the deployment, bookkeeping and maintenance for both experimenters and participants, while also allowing for better control over the actual code executed.

Experimenters, in short, need to implement the business logic of their experiment according to specific guidelines, and then submit it to OrganiCity for execution. This code is then distributed amongst OC volunteers, via the experimentation app on Android smartphones that have opted to participate in the experimentation process. Collected data are automatically gathered and stored by the OC backend. OC users have the ability, at all times, to stop or pause the execution of an experiment, or even opt out of the entire process. The final results are provided to experimenters in a fully anonymized format. Developing and coordinating an experiment typically requires between 30-100 lines of Android code.

In order to gather data, OC experimenters can use all the available sensor components available in current Android Smartphones of the OC Users, while also maintaining a certain degree of transparency, in order to allow OC Users to understand what kind of data they are reporting (i.e., contributing to OrganiCity). Such sensors include (but are not limited to) temperature, humidity, noise, walking steps, location and access to interfaces like WiFi, Bluetooth or NFC. A number of sensor components is already developed and available for experimenters to use, but additional ones can be implemented by experimenters based on their needs.

---

[4] http://ambientdynamix.org/

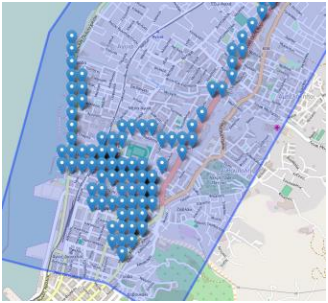

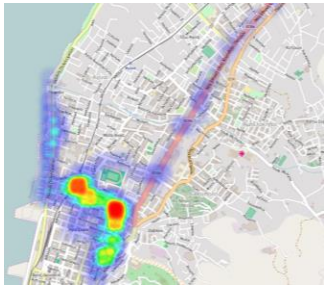

Figure 3: Experiment Monitoring view using points (top) and heatmaps (bottom). Through such views, experimenters can monitor the progress towards completing the experiments' crowdsensing goals, i.e., how are volunteers contributing readings in the designated areas and time periods.

Experimentation control in the context of the smartphone experimentation component refers to *the ability to control parameters of the execution of the experiment both before and during the deployment of such experiments*. This essentially translates to the ability to set certain restrictions on the experiment execution, when the experiments are initially set up:

> Restrictions on where the experiments are going to be executed. This means that we want to be able to determine distinct, multiple and areas of any shape, in the form of polygons.
> Restrictions on when the experiments are performed. Apart from the basic time period of the whole experiment, we want to be able, in the same way, to define discrete time periods of interest (i.e., there is little interest in values between 2-6 am).
> Restrictions on the minimum or maximum number of measurements for a specific spatio-temporal region. E.g., when measurements in an area have already reached a critical mass, the system could coordinate volunteers to other areas to achieve their goals as well.

The above restrictions help in a more homogeneous spatiotemporally execution of the experiments as not only is the geographical area of execution important, but the temporal plain is also taken into account. Apart from the ability to define these restrictions, the system provides feedback options and interfaces to monitor the progress/current state of execution of the experiments and the degree to which the constraints are fulfilled during the execution; e.g., the percentage of the requested readings already gathered.

**Experimentation Workflow**

After initial testing and development of the smartphone experimentation's code, the experimenter proceeds to the registration of the experiment on the OC Experimentation Management tool. To better illustrate each step of the process we hereby provide an example of an experiment performed simultaneously in 3 cities. The target of the experiment is to record WiFi access points around the London, Santander and Patras and use the data gathered to provide geolocation information to citizens.

To begin with, experimenters need to provide some *basic information* that describe the main goals of the experiment the usage intended for the data collected and a link to the outcomes of the experiment. All data provided here are available to volunteers, so they clearly understand what is executed on their phone and what data are collected and made available to the experimenter. In our case, we wrote a small text about how to participate in the experiment and created a static web page where the data collected are available for visitors to browse them.

*Spatial experimentation parameters*: The next step concerns the designation of experimentation areas in which the volunteers need to move. The experimenter freely designs polygons of interest without any limit on the number, size or location. Polygons can also overlap as the user can later on place restrictions on data collection that create different data characteristics. For example, an experimenter can select the whole city as a big polygon and then draw inside smaller areas of major interest or set different time constraints. In our case, we defined a set of polygons that depicted the main areas and suburbs of the three cities. For each city we created polygons of different detail levels (larger areas in London, smaller areas in Santander and a single polygon for Patras).

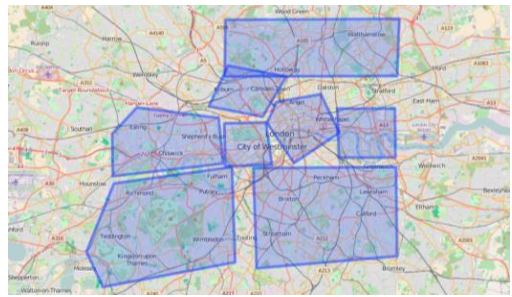

Figure 4: View of the polygons defined for London.

*Temporal experimentation parameters*: The next step concerns the definition of date, time and size constraints on the data acquisition for each polygon, as well as importance priorities. These parameters are used by the system to automatically guide volunteers to contribute more on more important regions or avoid areas that are already adequately explored. Similarly, we defined different time periods in the three cities as people tend to commute in different times of the day around the world.

*Selection of sensor plugins*: In this step, experimenters selects the sensor plugins this experiment requires. Sensor plugins can be already available in the system or uploaded as new Sensor plugin for this experiment. New Sensor plugins can also be selected to be public, so that other experimenters can use them in their own experiments as well. In our case, we reused the Location Sensor already available in the system and created our own WiFi scanner sensor plugin to scan for WiFi access points.

*Experiment plugin upload*: The final step is the upload of the Experiment's plugin code. This code will be validated by Organicity and then be made available to experimenters. After this step, the experiment is ready to be distributed to volunteers.

**Participating in an Experiment**

Volunteers are currently able to install the smartphone app through the OrganiCity website. They can at any time disable the experimentation process, using a "power-off" button that marks at all times the data collection process. Selecting, enabling and installing sensors and experiments can be easily performed by simply selecting a checkbox (Figure 1). Volunteers can select and register for a selected experiment. The Experiment is then downloaded in the background and execution starts on their phone. Should they need to enable more sensors, a warning message is displayed and they must return to the Sensors Tab to install the required sensors. An additional tab is also available with various statistics extracted from the data generated in real time (Figure 2). The statistics concern both the current experiment executed and the overall statistics for this specific volunteer.

| | |
|---|---|
| 3 | Cities |
| 14 | Participants |
| 13 | Regions |
| 5 | Experimentation Days |
| 7634 | Measurements |
| 40% | Avg Completion Rate |

Table 1: Statistics of the WiFi Scanner Experiment.

Completeness data are made available for each polygon defined. The data are presented either in summary tables (e.g., for the per hour completeness) or using map visualizations (Figure 3). The map visualizations are offered to OC experimenters in near-real time and generally aim at providing the experimenters with a better understanding of the movement patterns of their volunteers or the weak points of their experiment. In addition to that the data presented here can also be used by the experimenters to better incentivize and engage with volunteers. Upon the completion of an experiment, or even during the execution, OC experimenters can easily download the data produced in various formats (JSON, CSV etc.). The two formats available were selected as they can be used in conjunction with online visualization libraries.

In the WiFi Scanner experiment each volunteer contributed an average of 500 measurements adding up to a total of more than 7000 data points. Some of the points were gathered in areas of the cities outside the areas of interest defined but are kept in the system until the completion of the experiment in case the experimenter decides to change the limits of the polygons as new hot spots are identified inside the cities. Table 1 briefly presents a number of statistics on the experiment extracted by the system.

## Conclusions

We presented a system for organizing and monitoring crowdsensing campaigns over smartphones. Our system effectively allows for city-scale crowdsensing campaigns, also connecting to a smart city platform. We plan to continue our work and explore strategies for user incentives, develop additional plugins, and conduct campaigns with larger groups of volunteers.

## Acknowledgments
This work has been partially supported by the EU research project Organicity, under contract number H2020-645198.